\documentclass[%
 reprint,
 amsmath,amssymb,amsfonts
 aps,
prb,twocolumn
]{revtex4-2}
\usepackage[dvipdfmx]{graphicx}
\usepackage{graphicx}
\usepackage{bm}
\usepackage{txfonts}
\usepackage{color}
\usepackage{bm}
\usepackage[version=3]{mhchem}
\usepackage{mathtools}
\usepackage{newtxtext}
\usepackage[varg]{newtxmath}
\usepackage[normalem]{ulem}
\usepackage{lipsum}
\usepackage{here}
\usepackage{braket}
\usepackage{comment}
\usepackage[final]{hyperref}
\usepackage{txfonts}
\usepackage{color}
\usepackage[dvipdfmx]{graphicx}
\usepackage{here}
\usepackage{graphicx}
\usepackage{bm}
\usepackage{booktabs}
\usepackage{multirow}
\usepackage[mathlines]{lineno}

\usepackage{bm}
\usepackage{txfonts}
\usepackage{color}
\usepackage{color}
\usepackage{here}
\usepackage{dcolumn}
\usepackage{bm}
\usepackage[version=3]{mhchem}
\usepackage{mathtools}
\usepackage{newtxtext}
\usepackage[varg]{newtxmath}
\usepackage[normalem]{ulem}
\usepackage{lipsum}
\usepackage{braket}
\usepackage{comment}
\usepackage[final]{hyperref}

\begin{document}


\title{Magnetization and exchange-stiffness constants of Fe-Al-Si alloys at finite-temperatures: A first-principles study }

\author{Shogo Yamashita}\email{shogo.yamashita.q1@dc.tohoku.ac.jp} 
 \author {Akimasa Sakuma}

\affiliation{Department of Applied Physics, Tohoku University, Sendai 980-8579, Japan}%




\date{\today}
\begin{abstract} {We investigated the magnetic properties of Sendust (Fe-Al-Si) alloys not only {at} 0 K but also {at} finite-temperature{s} {by means of} the first-principles calculations assuming A2, B2, and DO3 structures. We confirmed that {the itinerant characteristics} of $3d$ electrons of Fe are not negligible for A2 and B2 structures and {a} significantly small exchange stiffness constant {exists} at zero-temperature {in} {a} B2 structure. However, the calculated Curie temperatures are in {the} same order for all structures; this indicates that {the} Curie temperature cannot be determined {only} by the exchange interactions  at zero-temperature in itinerant electron systems. Temperature dependence of the exchange interaction, namely spin configuration dependence, also might be important for determining it. In addition, {this property} might also {be} related to the unique behavior of the temperature dependence of the exchange stiffness constant for {the} B2 structure, which does not decrease monotonically as temperatures {increase, contrary} to the behavior expected from the Heisenberg model. {In addition}, we investigated composition dependence on the exchange stiffness constant at zero-temperature and confirmed that {the} substitution of Si {with} Al {could} improve the amplitude of the exchange stiffness constant at zero-temperature for all structures.} 
\end{abstract}
\maketitle
\section{Introduction} 
Magnetic materials are essential materials for sustaining {the wide range of }modern technology. For instance, hard magnets {with} strong magnetic anisotropy, such as $\textrm{Nd}_{2}$$\textrm{Fe}_{14}$$\textrm{B}$, are typical examples of magnetic materials {that} sustain modern society.
{Conversely}, soft magnetic materials {that} have weak magnetic anisotropy are also important materials for modern spintronics applications.
For instance, Sendust (Fe-Al-Si) alloy is a  soft magnetic material comparable with permalloy{s}. Sendust alloy {was} invented by Masumoto and Yamamoto in 1937\cite{Masumoto1937}, and {is now used in various} applications such as magnetic recording heads.
In addition, recently, Akamatsu et al\cite{Akamatsu2020,Akamatsu2022,Akamatsu2024}. {showed} that Sendust alloy is a promising candidate {for} the free layers of a magnetic tunnel junction (MTJ)-based magnetic sensor, which can detect external magnetic field{s} using the tunnel magnetoresistance (TMR) effect. 
{Generally, a high TMR ratio is desired to develop highly sensitive magnetic TMR sensors at room-temperature.}
However, the TMR ratio {is usually temperature dependent and } decreases as  temperatures {increase} due to, for instance, the thermal spin fluctuations.
Therefore, to sustain {a} high TMR ratio even at room temperature or above, it is essential to enhance the exchange stiffness constant, which represent{s the} robustness of the exchange couplings of a system, even at finite temperatures. \\
\ However, the theoretical description of {the} temperature dependence of the exchange stiffness constant is still controversial. Phenomenologically, the exchange stiffness constant $A$ can be defined as $f(\vec r)=A\sum_{\alpha}(\vec \nabla m_{\alpha}(\vec r) )^2$, where $f$ and $m$ are free energy density and magnetization density, respectively. The theoretical description of {the} temperature dependence of $A$ has been established, for instance, by Dyson\cite{Dyson1956} with the Heisenberg Hamiltonian. The recent attempts based on the localized spin model {have also been made}\cite{Atxitia2010,Moreno2016}. However, $3d$ electrons of transition metals are currently regarded as itinerant electrons. $3d$ electrons have dual properties, {that of} localized magnetic moments and hybridization effects, due to {their} itinerant {nature; moreover,} the magnetic moments and effective exchange interactions have environment dependence, for instance, spin-configurations dependence\cite{Small1984,Heine1990,Luchini1991,Mryasov1996}. Therefore, the localized spin model is not always suitable to describe itinerant magnets. Thus, we need to establish a theory {that} can capture both {the} localized and itinerant properties of $3d$ electrons {to describe the} temperature dependence of {the} exchange stiffness constant of itinerant magnets.  \\
\ From {this perspective}, many attempts to establish the spin-fluctuation theory of itinerant electron systems at finite temperatures have {already been}  made\cite{Moriyabook}. Among them, a convenient theoretical way to {describe} finite temperature magnetic properties {by} describing spin-fluctuations of itinerant magnets is the functional integral method combined with the {concept of the} local moment disorder (DLM) and coherent potential approximations (CPA)\cite{Cyrot,Hubbard1,Hubbard2,Hasegawa1,Hasegawa2,Hasegawa3}. {This method was originally} proposed in the single band Hubbard model and {was} translated to {a} first-principles calculation scheme to investigate finite-temperature magnetic properties based on realistic electronic structures\cite{Oguchi,Pindor,Staunton3,Gyorrfy,StauntonPRL1992}. Although this scheme is based on static and single-site approximations, recently, it has been applied to {calculate}  the temperature dependence of magnetic anisotropy\cite{StauntonPRL,StauntonPRB,Deak1,Matsumoto,Yamashita2022,Juba1,Yamashita2023,woodgate2023}, Gilbert damping constant\cite{Hiramatsu1,Hiramatsu2023}, and transport properties\cite{Sakuma2022}, and {the}  non-localized spin model like behavior of finite-temperature magnetic properties have been revealed\cite{Yamashita2023}. In addition to these properties, very recently, Sakuma\cite{Sakuma2024} proposed the scheme to calculate the temperature dependence of the exchange stiffness constants of itinerant magnets based on the DLM-CPA scheme. {This has enabled the investigation of} the temperature dependence of the exchange stiffness constant of itinerant magnets without assuming localized {spin model} for $3d$ electrons in metallic systems. \\
\ In this study, we aim to investigate the magnetic properties of Fe-Al-Si alloys not only at 0K but also at finite temperatures;  in particular, {we aim to investigate the} temperature dependence of the exchange stiffness constant by {using} the finite-temperature itinerant electrons theory based on first-principles calculations with the DLM-CPA scheme.
\section{Method} 
\begin{flushleft} 
\begin{figure*}[h]
\begin{center}
\includegraphics[clip,width=13.5cm]{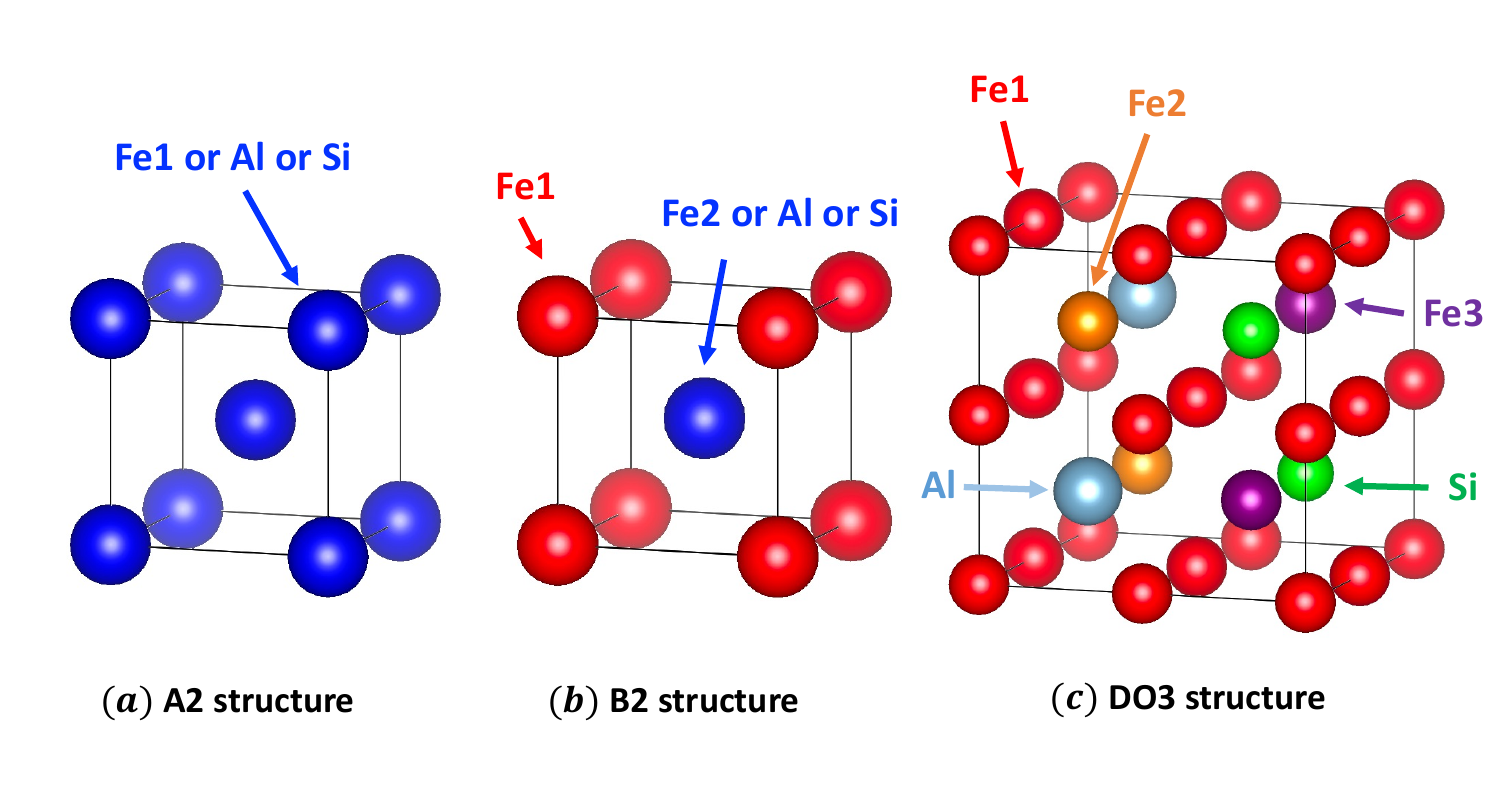}
\end{center}
\caption{Crystal structures of Fe-Al-Si alloy $(\textrm{Fe}_{75}\textrm{Al}_{12.5}\textrm{Si}_{12.5})$ assumed in this study.}
\label{Crystalstructure}
\end{figure*} 
\end{flushleft}

\ We use the functional integral approach together with the local moment disorder picture to investigate {the} finite-temperature magnetic properties of Fe-Al-Si based on first-principles calculations. We treat the fluctuation of the local moments with the adiabatic approximation due to the slow time scale of the spin fluctuation compared with the motion of electron hopping.
The partition function $Z$ and the free energy $F$ of the system are given as follows:
\begin{align}
Z=\int \left(\prod_{i} \textrm{d} {\bf{e}}_{i} \right) e^{-\Omega(T,\lbrace{\bf {e}}\rbrace)/k_{\textrm{B}}T},
\end{align}
\begin{align}
F=\langle \Omega(T,\lbrace{\bf {e}}\rbrace)\rangle_{\omega(T,\lbrace{\bf {e}}\rbrace)}+k_{\textrm{B}}T \langle \textrm{ln} \ \omega(T,\lbrace{\bf {e}}\rbrace)\rangle_{\omega(T,\lbrace{\bf {e}}\rbrace)},
\label{FreeE}
\end{align}
where $\Omega(T,\lbrace{\bf {e}}\rbrace)$, $\omega(T,\lbrace{\bf {e}}\rbrace)$, and $ {\bf{e}}_{i}$ are a thermodynamic potential, a  distribution of spin orientation, and an orientation vector of {the} local spin moment at site $i$. $k_{\textrm{B}}$ is a Boltzmann constant.  {Once we obtain $\Omega(T,\lbrace{\bf {e}}\rbrace)$, $\omega(T,\lbrace{\bf {e}}\rbrace)$ is given as follows:
\begin{align}
 \omega(T,\lbrace{\bf {e}}\rbrace)=\textrm{exp} \left(-\Omega(T,\lbrace{\bf {e}}\rbrace)/k_{\textrm{B}}T \right)/Z.
\end{align}
} Here, we neglect the longitudinal fluctuations of local moments. {Thus, the integration with respect to ${\bf {e}}$ is only for the freedom of directions.}
$\langle A \rangle_{\omega(T,\lbrace{\bf {e}}\rbrace)}$ is given as {follows:} 
\begin{align}
\langle A(\lbrace{\bf {e}}\rbrace) \rangle_{\omega(T,\lbrace{\bf {e}}\rbrace)}=\int \prod_{i} \textrm{d} {\bf{e}}_{i}  \ \omega(T,\lbrace{\bf {e}}\rbrace)\ A(\lbrace{\bf {e}}\rbrace).
\end{align}
In this study, we use the tight-binding linearized muffin-tin orbital (TB-LMTO) method\cite{Sakuma2000,Andersen,Skriver,Turek,Kudrnovsky} to evaluate $\Omega(T,\lbrace{\bf {e}}\rbrace)$, which determines spin orientation distribution corresponding to the spin-transverse fluctuations at $T$. 
The Green function, including the spin-transverse fluctuation of the system in the TB-LMTO method, is given as follows:
 \begin{align}
G_{ij}(z;{\lbrace{\bf {e}}\rbrace})&=(z-H_{{\textrm{TB-LMTO}}}(\lbrace{\bf {e}}\rbrace))_{ij}^{-1} \nonumber \\
&=\lambda_{i}^{\alpha}(z;{\lbrace{\bf {e}}\rbrace})\delta_{ij}+\mu^{\alpha}_{i} (z;{\lbrace{\bf {e}}\rbrace}) g_{ij}^{\alpha} (z;{\lbrace{\bf {e}}\rbrace}) \bar \mu_{j}^{\alpha}(z;{\lbrace{\bf {e}}\rbrace}), 
\end{align}
 \begin{align}
g^{\alpha}_{ij}(z;{\lbrace{\bf {e}}\rbrace})=\left[(\tilde P^{\alpha}(z; {\lbrace{\bf {e}}\rbrace})-S^{\alpha})^{-1}\right]_{ij}, 
\end{align}
where $\lambda$ and $\mu$ are 
\begin{align}
\lambda^{\alpha}_{i}(z; {\bf{ e}}_{i} )=({\tilde \Delta_{i}}({\bf{e}}_{i}))^{-1/2}(1+(\tilde\gamma_{i}({\bf e}_{i})-\alpha)\tilde{P}_{i}^{\gamma}(z;{\bf{e}}_{i}))({\tilde\Delta_{i}}({\bf{e}}_{i}))^{-1/2},
\end{align}
\begin{align}
\mu^{\alpha}_{i}(z; {\bf{ e}}_{i} )=({\tilde\Delta_{i}}({\bf{e}}_{i}))^{-1/2}(\tilde {P}_{i}^{\gamma}(z;{\bf{e}}_{i}))^{-1}\tilde P^{\alpha}_{i}(z; {\bf e}_{i}),
\end{align}
\begin{align}
\bar \mu^{\alpha}_{i}(z;  {\bf{ e}}_{i} )=\tilde P^{\alpha}_{i}(z; {\bf e}_{i})(\tilde{P}_{i}^{\gamma}(z;{\bf{e}}_{i}))^{-1}({\tilde\Delta_{i}}({\bf{e}}_{i}))^{-1/2},
\end{align}
\begin{align}
P^{\alpha}_{i}(z)=P^{\gamma}_{i}(z)\lbrace1-[\alpha-\gamma_{i}]P^{\gamma}_{i}(z)\rbrace^{-1},
\label{Trans}
\end{align}
\begin{align}
P^{\gamma}_{i}(z)=(\Delta_{i})^{-1/2}[z-C_{i}](\Delta_{i})^{-1/2}.
\label{PG}
\end{align}
{Here,}  $\gamma_{i}$, $\Delta_{i}$, and $C_{i}$ are called potential parameters. {$S^{\alpha}$ is given as $S(1-\alpha S)^{-1}$. $S$ is a bare structure constant.}
{From here, to {utilize} the CPA, we use the $\beta$-representation, regarded as {a} maximum localized representation, instead of general $\alpha$ representations to express the Green function.}
{$\beta$ values {have been summarized in earlier studies} \cite{Sakuma2022,Turek,Kudrnovsky}. }
The Hamiltonian $H_{{\textrm{TB-LMTO}}}$ is given as follows:
{ \begin{align}
&H_{{\textrm{TB-LMTO}}}(\lbrace{\bf {e}}\rbrace)\nonumber \\ &=\tilde C(\lbrace{\bf {e}}\rbrace)+\tilde\Delta^{1/2} (\lbrace{\bf {e}}\rbrace)S (1-\tilde\gamma(\lbrace{\bf {e}}\rbrace)  S)^{-1}\tilde \Delta^{1/2}(\lbrace{\bf {e}}\rbrace).
\end{align}}
In this study, we defined $\tilde A(\lbrace{\bf {e}}\rbrace)$ as follows:
\begin{align}
\tilde A(\lbrace{\bf {e}}\rbrace)= U^{\dag} (\lbrace{\bf {e}}\rbrace)A  U(\lbrace{\bf {e}}\rbrace),
\end{align}
where $U$ is a SU(2) rotation matrix.\\
 \  The thermodynamic potential of {the} electronic part $\bar \Omega$ is also expressed as follows:
 \begin{align}
&\bar \Omega(T,\lbrace{\bf {e}}\rbrace)\nonumber \\ 
&=\frac{1}{\pi} \int \textrm{d} \epsilon f(\epsilon,T,\mu) \int_{-\infty}^{\epsilon} \textrm{d} E \ \textrm{ImTr} \ G (z;\lbrace {\bf{ e}}\rbrace)\nonumber
 \\ &=-\frac{1}{\pi} \int \textrm{d} \epsilon f(\epsilon,T,\mu)\textrm{Im} \left[\textrm{Tr}\ \textrm{log} \ \lambda^{\beta}(\epsilon^{+};\lbrace {\bf{ e}}\rbrace) + \textrm{Tr}\ \textrm{log} \ g^{\beta}(\epsilon^{+};\lbrace {\bf{ e}}\rbrace) \right],
 \end{align}
 where $f$ and $\mu$ are the Fermi--Dirac function and chemical potential, respectively; here we dropped the double counting part of {the} exchange-correlation energy part relying on the concept of the magnetic force theorem\cite{Lichtenstein1987} and {evaluated} $\omega(T,\lbrace{\bf {e}}\rbrace)$ with only {the} electronic part of $\Omega(T,\lbrace{\bf {e}}\rbrace)$.\ In {the} actual evaluation of $\bar \Omega(T,\lbrace{\bf {e}}\rbrace)$, we need to {perform}  the functional integral with respect to various spin configurations composed {of}  multi spins.
 {This is generally impractical and an approximation needs to be made. } 
 To obtain $\bar \Omega(T,\lbrace{\bf {e}}\rbrace)$, we use the coherent potential approximation together with the single-site approximation.
 In this approximation, it will also be decomposed as follows:
 \begin{align}
\bar \Omega(T,\lbrace{\bf {e}}\rbrace)=\bar\Omega_{0}+\Delta \bar\Omega(T,{\bf {e}}_{i}),
 \end{align}
 and each part is also given as {follows:} 
  \begin{align}
 \bar \Omega_{0}=-\frac{1}{\pi} \int \textrm{d} \epsilon f(\epsilon,T,\mu)\textrm{Im} \left[\textrm{Tr}\ \textrm{log} \ \bar g^{\beta}(\epsilon^{+}) \right],
 \end{align}
  \begin{align}
 \Delta \bar\Omega(T,{\bf {e}}_{i})&=-\frac{1}{\pi} \int \textrm{d} \epsilon f(\epsilon,T,\mu)\textrm{Im} \biggl[\textrm{Tr}\ \textrm{log} \lambda^{\beta}(\epsilon^{+}) \nonumber \\&- \textrm{Tr}\ \textrm{log} (1+\Delta P(z;\lbrace{\bf e}\rbrace)\bar g^{\beta}(\epsilon^{+})  )\biggr],
 \end{align}
 \begin{align}
\Delta P(z;\lbrace{\bf e}\rbrace)=\tilde P^{\beta}(z; \lbrace{\bf e}\rbrace)-\bar P(z).
\label{expandon}
\end{align}
Here, we define $\bar g^{\beta}$ as follows:
\begin{align}
\bar g^{\beta}(z)= (\bar P(z)-S^{\beta})^{-1},
\label{CPAG}
\end{align}
where $S^{\beta}$ is also given as $S(1-\beta S)^{-1}$. To obtain $\bar P(z)$, we use {the so-called} CPA condition.
The CPA condition is given as {follows:}
\begin{align}
{\int \textrm{d} {{\bf e}}_{i} \ \omega_{i}(T,{\bf{e}}_{i})  \left[1+ \Delta P_{i}(z;{\bf e}_{i})\bar g^{\beta}_{ii}(z)\right]^{-1} \Delta P_{i}(z;{\bf e}_{i})=0.}
\label{CPAcon}
\end{align}
Once we have converged $\omega_{i}(T,{\bf {e}}_{i})$, we can obtain $\bar P$ in {a} self-consistent manner using Eqs. \eqref{expandon}, \eqref{CPAG}, and \eqref{CPAcon}.
Therefore, we can obtain $\omega_{i}(T,{\bf {e}}_{i})$ as follows:
 \begin{align}
&\omega_{i}(T,{\bf{e}}_{i}) \nonumber \\&=\textrm{exp} \left(-\Delta\bar \Omega_{i}(T,{\bf{e}}_{i})/k_{\textrm{B}}T \right)/ \int \textrm{d} {\bf e}_{i} \textrm{exp} \left(-\Delta\bar\Omega_{i} (T,{\bf{e'}}_{i})/k_{\textrm{B}}T\right).
 \end{align}
 In this approximation, $\omega({T,\lbrace{\bf {e}}\rbrace})$ {is} decomposed to {the} simple product of probability at each site $i$ as $\omega(T,\lbrace{\bf {e}}\rbrace)=\prod_{i}\omega_{i}(T,{\bf {e}}_{i})$. \\
 \ To calculate the exchange stiffness constant, we express a spin-spiral magnetic structure\cite{Sandratskii1986,KublerBook,Mryasov1992,Uhl1992,Kubler2006}.
 In the TB-LMTO scheme, {the} structure constant $\tilde S$ to express spin-spiral states is given as follows\cite{Yamashita2023APEX,Yamashita2024}:
 \begin{align}
 \tilde S_{ij}(\vec k,\vec q)= U_i \begin{pmatrix}
 S_{ij}(\vec k-\frac{\vec q}{2}) & 0 \\
0 & S_{ij}(\vec k+\frac{\vec q}{2}) \\
\end{pmatrix}U^{\dag}_j{,}
\label{SSP}
 \end{align}
 {where $i$ and $j$ stand for sites of atoms in a primitive unit cell, and we redefine $U$ at site $i$ as follows:}
 \begin{align}
 U_{i}=\frac{1}{\sqrt{2}}
 \begin{pmatrix}
   e^{i(\frac{\vec q}{2}\cdot \vec r_{i})}& e^{-i(\frac{\vec q}{2}\cdot \vec r_{i})} \\
   -e^{i(\frac{\vec q}{2}\cdot \vec r_{i})}& e^{-i(\frac{\vec q}{2}\cdot \vec r_{i})}  \\
\end{pmatrix},
 \end{align}
 where $\vec r_{i}$ denotes atomic positions in a primitive unit cell. 
{We also rely on the concept of the force theorem to obtain the free energy difference. }
{Neglecting the contribution of the second term {in} Eq. \eqref{FreeE} (entropy of the spin configuration) as well as the double counting term,} the free energy difference is given as follows:
 \begin{align}
\Delta F(\vec q, T)\sim \langle \bar \Omega(\vec q,T)\rangle- \langle \bar \Omega(\vec q=0,T )\rangle,
\end{align}
\begin{align}
\langle \bar\Omega(\vec q,T) \rangle&=\sum_{i} \int \textrm{d} {\bf e}_{i} \  \omega_{i}({\bf{e}}_{i},T) \ \bar \Omega_i(T,\vec q,{\bf e}_{i}),
\label{ENESOC}
\end{align}
\begin{align}
& \bar \Omega_i(T,\vec q,{\bf e}_{i})=-\frac{1}{\pi} \textrm{Im}\textrm{Tr}_{L\sigma} \int _{-\infty}^{\infty}\textrm{d}  \epsilon \ \epsilon f(\epsilon,T,\mu) G_{ii}(\epsilon+i\delta,\vec q,{\bf e}_{i}) \nonumber \\ &-\frac{k_{\textrm{B}}T}{\pi}\textrm{Im}\textrm{Tr}_{L\sigma}\int _{-\infty}^{\infty}\textrm{d}  \epsilon \ G_{ii}(\epsilon+i\delta,\vec q,{\bf e}_{i}) \nonumber \\ &\times \left(  f(\epsilon,T,\mu) \textrm{log} f(\epsilon,T,\mu)+(1- f(\epsilon,T,\mu)) \textrm{log} (1- f(\epsilon,T,\mu))  \right),
\label{Omega}
\end{align}
{
\begin{align}
&G_{ii}(z,\vec q,{\bf e}_{i})\nonumber \\ &=\lambda^{\beta}_{i}(z,{\bf {e}}_{i})+\mu^{\beta}_{i}(z,{\bf {e}}_{i})   \bar g^{\beta}_{ii}(z,\vec q) (1+\Delta P_{i}(z,{\bf {e}}_{i})\bar g^{\beta}_{ii}(z,\vec q))^{-1}  \bar \mu^{\beta}_{i}(z,{\bf {e}}_{i}),
\end{align}}
{
\begin{align}
  \bar g^{\beta}_{ii}(z,\vec q)=\left(\frac{1}{N}\sum_{\vec k} \left(\bar P(z)-\tilde S^{\beta}(\vec k,\vec q)\right)^{-1}\right)_{ii},
 \end{align}}
 {where $N$ is a number of $\vec k$-points.}
Using this free energy difference, the exchange stiffness constant $A_{\eta}(T)$ is given as follows: 
\begin{align}
A_{\eta}(T)=\frac{1}{2V} \frac{d^2 \Delta F (\vec q, T)}{ d q_{\eta}^2},
\label{exA}
\end{align}
where $V$ is {the} volume of the system, and $\eta$ is the direction of the spin-spiral.
In this study, we fix $\eta$ {to be in the} [001] direction $(q_{\eta}=q_{\textrm{z}})$.
We show the assumed crystal structures of Fe-Al-Si in Fig. \ref{Crystalstructure}.
In this study, we set the composition of these elements to Fe 75\%, Al 12.5\%, and Si 12.5\% {to calculate the } finite-temperature magnetic properties.
The lattice constant $a$ is set to $2.884$ $\textrm{\AA}$  for  {the} A2 and B2 structure and $5.688$ $\textrm{\AA}$  for {the} DO3 structure.
We used the local spin density approximation for the exchange correlation potential.
\section{Results and discussions} 
 \begin{flushleft} 
\begin{figure}[h]
\begin{center}
\includegraphics[clip,width=8.5cm]{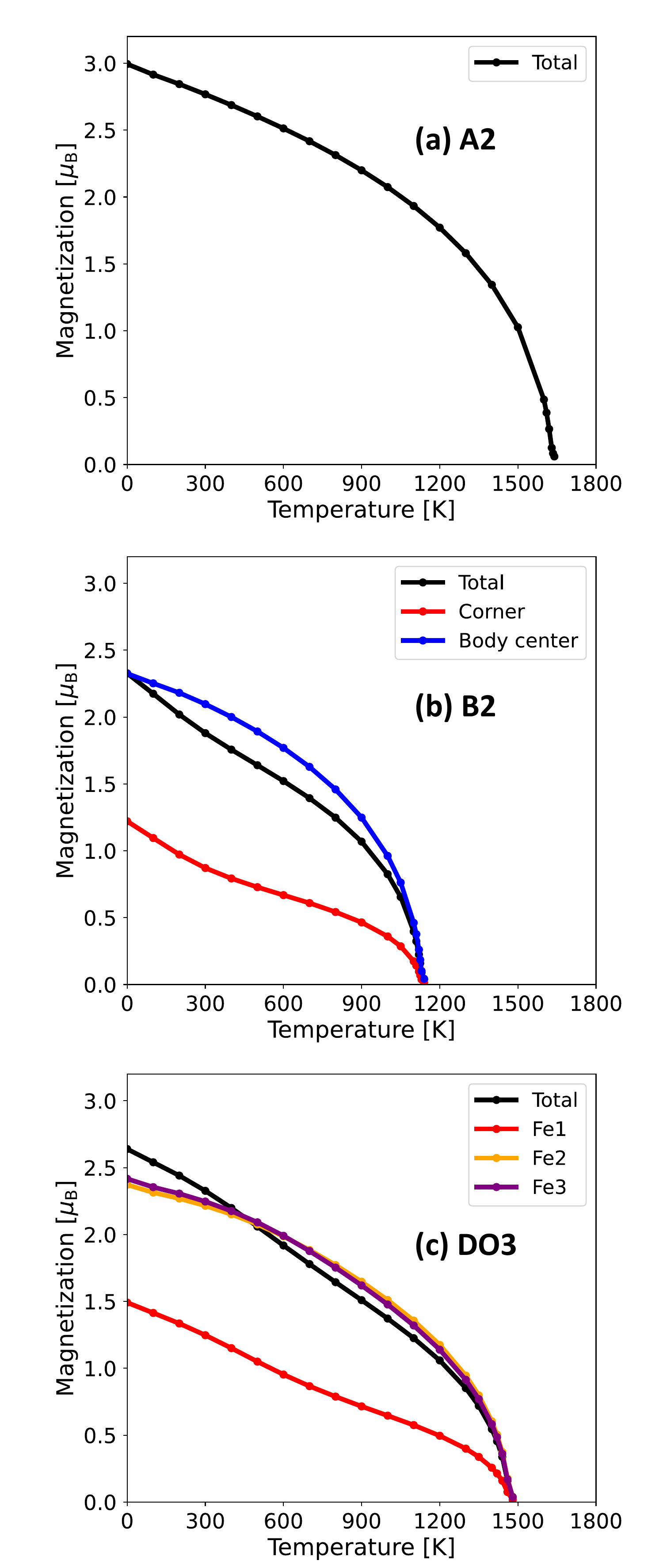}
\end{center}
\caption{Calculated temperature dependences of magnetization for (a) A2, (b) B2 and (c) DO3. The estimated Curie temperatures are $1160$ K, $1640$ K, and $1480$ K for {the}  A2, B2 and DO3 structures, respectively. For comparison, we scaled the total magnetization of {the} DO3 structure {by} 1/8. Temperature dependences of Fe sites resolved magnetization are also shown in (b) and (c).}
\label{MT}
\end{figure} 
\end{flushleft}
\ Initially, we summarize {the} calculated zero-temperature magnetic properties of Fe-Al-Si alloys.
{The} calculated exchange stiffness constants and magnetic moments of each structure are summarized in Table \ref{MA}.
From this table, we can see that the exchange stiffness constant of {the} B2 structure is significantly small (about one-tenth) compared with that of other structures.
{We will conduct further investigation regarding this in a latter section. }
In the B2 structure, the magnetic moment of the Fe1 site is significantly small compared with that of the other site. {A similar tendency is confirmed in the} Fe1 site in {the} DO3 structure. This reduction of magnetic moments may {be attributed} to the hybridization effects of {the} $3d$ orbitals of {Fe1 sites with the 2p orbitals of Al or Si. The interatomic distance between Fe1 and Al or Si is $\frac{\sqrt{3}a}{2}$, which is smaller than $a$, corresponding to the distance between Fe2 and Al or Si.} To investigate it, we also performed the self-consistent conventional DLM calculations, which include only spin up and down states for A2 and B2 structures, and calculate {the} converged magnetic moments of each site. {The} calculated results are also summarized in Table \ref{MA}. From these results, {the} significant reduction of {the} magnetic moments of {the}  Fe1 site of {the} B2 structure is confirmed; this reflects {the} itinerant picture of $d$ electrons in Fe1 sites, which have a strong dependence on {the}  environment surrounding the atoms. 
This situation can be {attributed to} the electronic state of Fe in {the} Fe1 site {being}  close to that of fcc Co and fcc Ni {observed in several studies} \cite{Uhl1996,Rosengaard1997,Halilov1998,Ruban2007} { and it is  owing to the hybridization with the $2p$ orbitals of light elements such as N, B, C and Al or Si in this case. {Conversely}, the magnetic moments of Fe2 sites are larger than that of bcc-Fe; this can be considered a compensation effect due to the hybridization with the Fe1 $3d$ orbitals, whose exchange splitting is reduced. In the A2 structure, the Fe moments go to an intermediate value because the two effects on Fe1 and Fe2 are {weakened}.}
\begin{table}
    \centering
    \caption{Calculated exchange stiffness constants and magnetic moments at each site of Fe-Al-Si alloys with each structure at ferromagnetic and the DLM states.
 }\begin{ruledtabular}
    \begin{tabular}{ccccc}
        System &$A$ [meV/\AA] & $M_{\textrm{Fe1}}$ [$\mu_{\textrm{B}}$]   & $M_{\textrm{Fe2}}$ [$\mu_{\textrm{B}}$]  & $M_{\textrm{Fe3}}$ [$\mu_{\textrm{B}}$]  \\
    \hline  
        A2    &    10.93     & 2.029        & -        & -              \\
        B2     &   0.938     & 1.220         & 2.325       & -                \\
        DO3    &    8.480    & 1.491        & 2.372        & 2.417                \\
   \hline   
        DLM A2    & -        & 1.766        & -        & -                \\
        DLM B2    & -        & 0.015        & 2.438        & -              \\
   \end{tabular}
    \label{MA}
    \end{ruledtabular}
\end{table}
\\
\\
\ Next, we investigate {the} finite temperature magnetic properties of Fe-Al-Si alloys. At first, we show the temperature dependences of the magnetization of Fe-Al-Si alloys for each structure in Fig. \ref{MT}. The result for {the}  A2 structure, the shape of {the} total magnetization curve is similar to the Langevin function, which can be expected from the single-site approximation. However, other results do not follow this trend. 
To examine {the reason for} this situation, we focus on {the} temperature dependence of site-resolved magnetization for B2 and DO3 structures.
For the B2 and DO3 structures, {the} behavior of the temperature dependence of magnetization at Fe1 sites in {the} B2 structure and Fe1 sites in {the}  DO3 structure in Fig. \ref{Crystalstructure} (c) deviates {from} the Langevin function-like behavior. In addition, the magnetic moments of these sites at 0 K are smaller than that of other sites; the exchange coupling energy $J_{0}$\cite{Lichtenstein1987} is also smaller than that of other site.
{As discussed earlier, this may be due to the fact that the} magnetic moments on these Fe atoms are easily influenced by environment {wherein, they are} surrounded by other atoms and {the} localized {characteristics} is weakened. 
{Owing to} above reasons, {the} magnetic moments of these sites are sustained by the exchange field from other Fe atoms, {not by the intra-atomic Coulomb interaction (Hund coupling)} at high temperatures.
\ {The} calculated Curie temperatures are $1640$ K, $1160$ K, and $1480$ K for {the}  A2, B2, and DO3 structures, respectively. Experimentally, it is reported {to be} 733 K\cite{Wang2008} for $\textrm{Fe}_{85}\textrm{Al}_{5.4}\textrm{Si}_{9.6}$ (wt. \%) ({$\textrm{Fe}_{73.7}\textrm{Al}_{9.7}\textrm{Si}_{16.6}$ (at. \%)}). Although the composition is different from our calculation models, our results overestimate the Curie temperature compared with the experimental result. {It is noted that}{ as shown by Hiramatsu et al.\cite{Hiramatsu1}, the Curie temperature of bcc-Fe is approximately 2000 K which is {approximately} two times larger than the experimental value. Judging from the present data combined with {that of} the of bcc-Fe, we see that the family of alloys related to bcc-Fe have a trend to exhibit high Curie temperature. {Considering this}, the lower $T_{\textrm{c}}$'s of the Fe-Al-Si alloys compared to the $T_{\textrm{c}}$ of bcc-Fe may be reasonable since a part of Fe atoms are substituted by non-magnetic elements such as Al and Si in these alloys.}
{It is worth mentioning that the reduction of the Curie temperature for bcc-Fe due to the phonon excitation has been reported theoretically\cite{Tanaka2020}. 
In addition, the reduction of the local moment induced by phonon excitation has also been reported by Heine et al.\cite{Heine2020} for bcc-Fe. 
Our case is close to the situation of bcc-Fe, thus it might affect our results. 
However, these couplings between spin disorders and phonon excitations are not taken into account in our study. Moreover, we used the force theorem approach and the potential parameters are fixed to that of the ferromagnetic ground states.
 These treatments also might affect our result, in particular for the results of B2 structure.}
 \begin{flushleft} 
\begin{figure}[h]
\begin{center}
\includegraphics[clip,width=8.5cm]{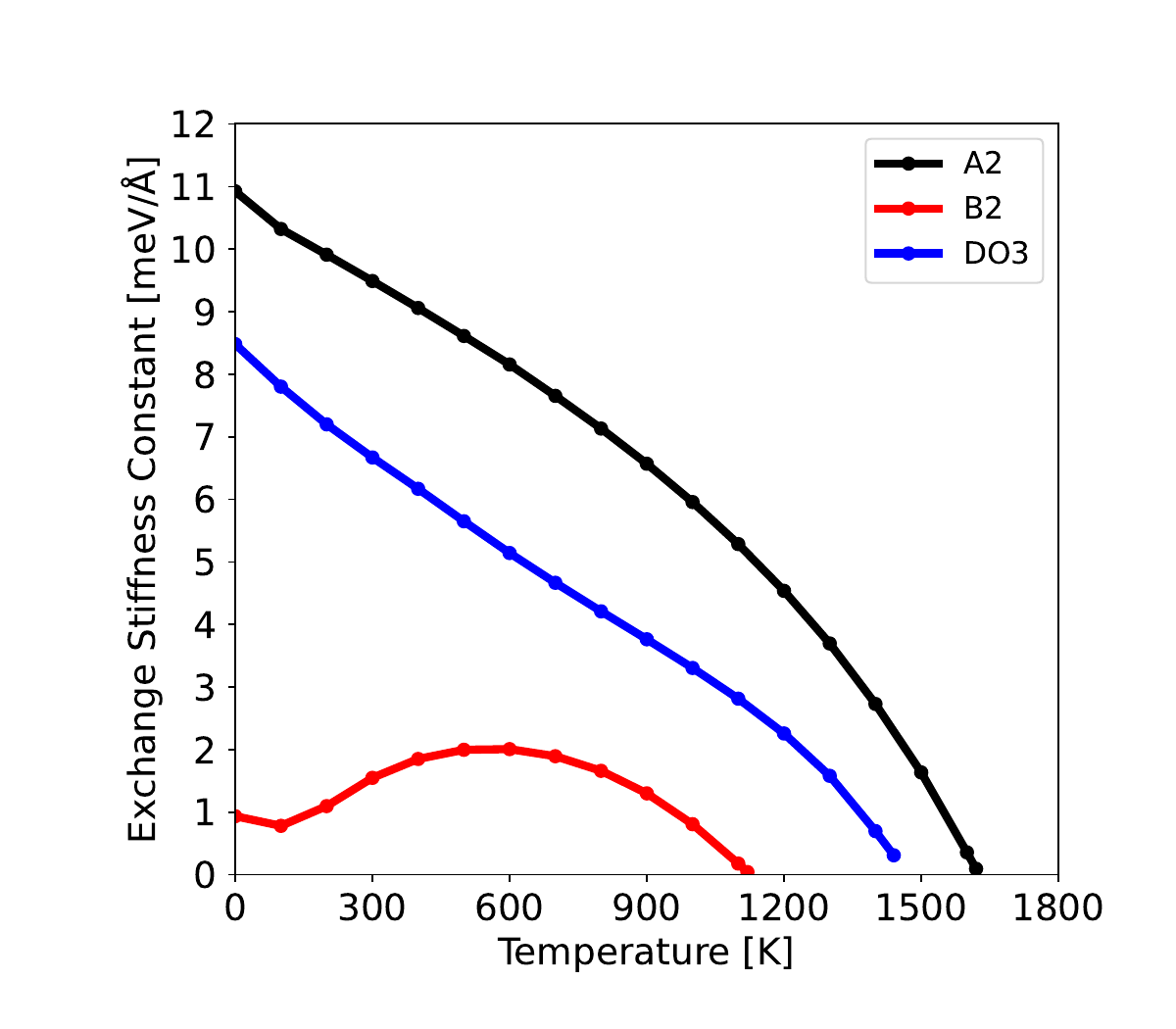}
\end{center}
\caption{Calculated temperature dependence of the exchange stiffness constants of Fe-Al-Si alloy with A2, B2, and DO3 structure. 
}
\label{AT}
\end{figure} 
\end{flushleft}
\ {{Moreover}, apart from the values of $T_{\textrm{c}}$ themselves, {the fact that}  the relative values of $T_{\textrm{c}}$ of these alloys are not proportional to the values of $A$ at $T=0$ as shown in Table \ref{MA}; that is,  the $A$ of B2 structure is extremely smaller compared to those of A2 and DO3 structures, while the $T_{\textrm{c}}$ of these structures are in the same order {is puzzling}. These results may indicate that {the} Curie temperatures of itinerant electron systems are not always determined by the exchange interaction at $T=0$ but are governed by the temperature variation of the exchange constants. To examine this feature, we calculated the temperature dependence of $A$ of these structures.} 
\\ \ {Figure \ref{AT} shows the calculated results of $A(T)$. The $A(T)$ of A2 and DO3 structures {was found to} exhibit monotonic decreasing behavior with increasing $T$, as seen in the measured spin stiffness constant $D(T)$ of bcc-Fe\cite{Shirane1968}. If one adopts a simple spin model such as the Heisenberg model, where the constant exchange interaction is given, $A(T)$ may decrease with increasing $T$. Actually, according to Dyson\cite{Dyson1956}, $A(T)$ behaves proportional to $1-(T/T_{\textrm{c}})^{\frac{5}{2}}$ in the low-temperature region owing to the magnon--magnon scattering. {However, as seen in Table \ref{MA}, the localized spin picture may not be appropriate for these alloys.} }{Conversely}, $A(T)$ of {the} B2 structure reveals a peak at around 600 K. 
 If one focuses on the behavior of $A(T)$ for $T>700$ K, the resultant $T_{\textrm{c}}$ of {the} B2 structure seems reasonable. 
 It should be emphasized here that, in metallic systems, the exchange interaction itself can vary depending on the spin configuration at each temperature, as {reported} for basic magnetic materials\cite{Heine1990}. {This} has been confirmed in Heusler alloys\cite{Khmelevskyi2022} and {even for a metallic $4f$ electron system.\cite{Khmelevskyi2007}}  {Although our calculations are  based on the force theorem, not self-consistent calculations performed in previous works to investigate configuration dependence of the exchange interactions, this effect might seem to be key to explain our result.} 
 At this stage, however, {why} this particular behavior {occur} only in the B2 structure {is not revealed exactly}. 

 \begin{flushleft} 
\begin{figure}[h]
\begin{center}
\includegraphics[clip,width=8.5cm]{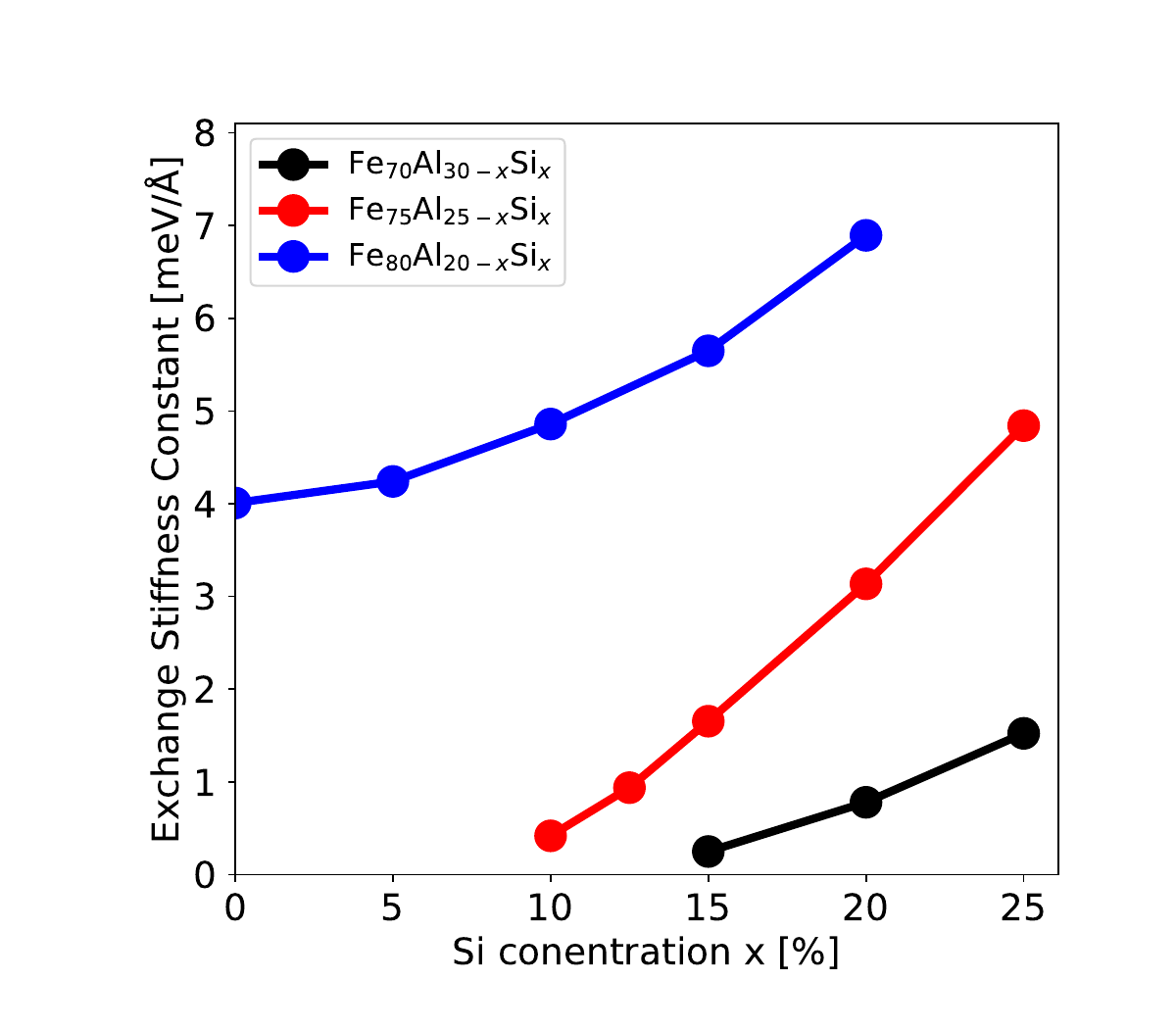}
\end{center}
\caption{Si concentration dependence of the exchange stiffness constant for {the} B2 structure at 0 K. The concentration of Fe {increases} from 70\% to 80\%. }
\label{AB2dep}
\end{figure} 
\end{flushleft}
{ \ To investigate how the Al and Si atoms affect the $A$ of {the} B2 structure, we examine the Si concentration dependence of $A$ at $T=0$. The results are shown in Fig. \ref{AB2dep}  by varying {the} Fe concentration from 70 to 80 atomic percent. The results clearly indicate that $A$ is significantly sensitive to the Si concentration; that is, $A$ decreases monotonically with decreasing Si concentration or increasing Al concentration.}
{Notably,} if the Si concentration is below 10\% for $\textrm{Fe}_{75}\textrm{Al}_{25-x}\textrm{Si}_{x}$ and 15\% for $\textrm{Fe}_{70}\textrm{Al}_{30-x}\textrm{Si}_{x}$, the ferromagnetic state becomes unstable, and the exchange stiffness constant cannot be defined. {We consider that this is attributed to the  itineracy of $3d$ electrons due to the strong hybridization between the Fe $3d$ orbitals and Al $2p$ orbitals, whose radius is larger than those of Si atoms.}
This result may indicate that {the}  concentration of {the} Al atom is a key of {the} small exchange stiffness constant of {the}  Fe-Al-Si alloy with {the} B2 structure.
 \begin{flushleft} 
\begin{figure}[h]
\begin{center}
\includegraphics[clip,width=8.5cm]{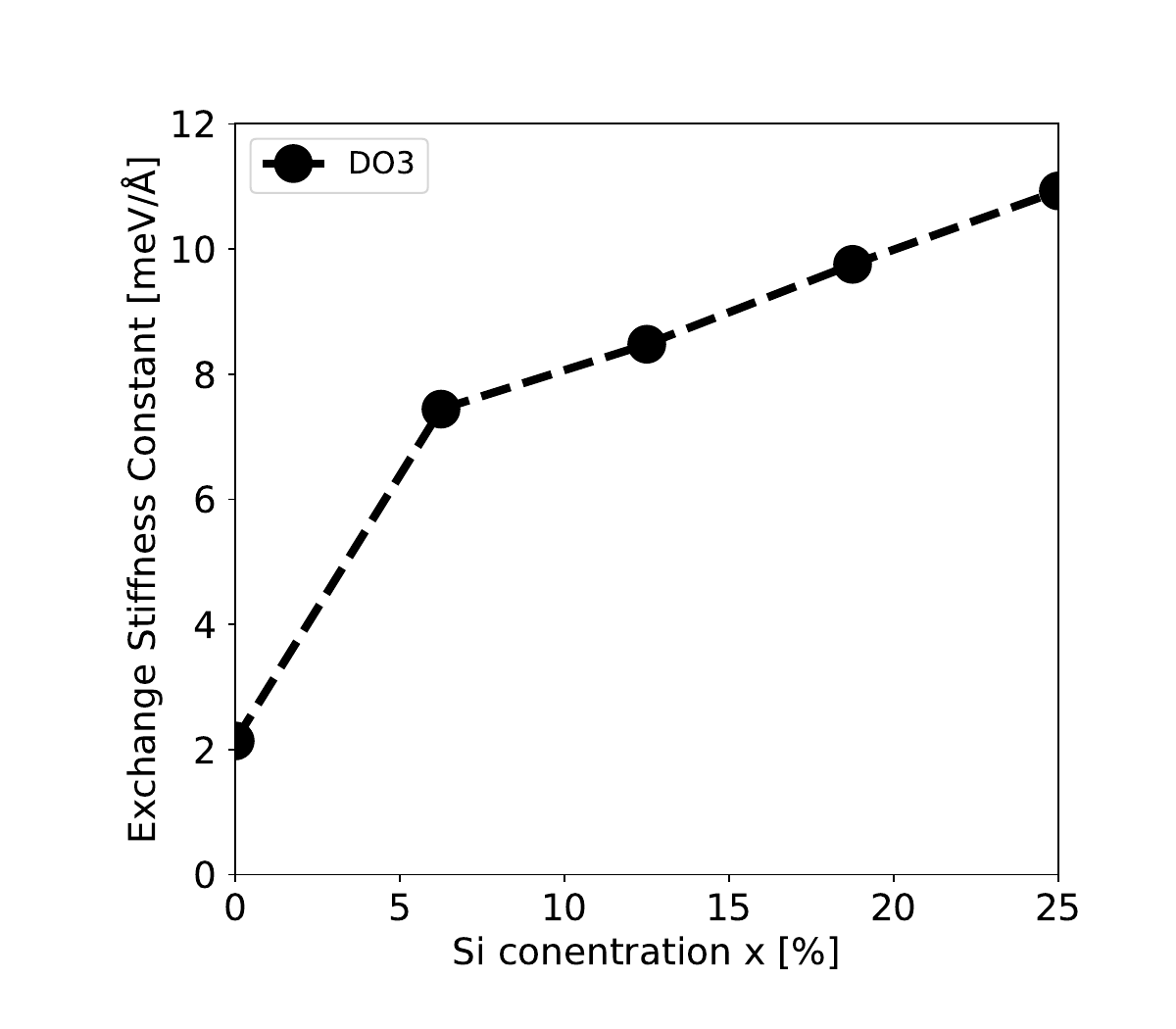}
\end{center}
\caption{Si concentration dependence of the exchange stiffness constant for {the} DO3 structure at 0 K.  }
\label{ADO3dep}
\end{figure} 
\end{flushleft}
\ {To {investigate} how the Al atoms affect the exchange stiffness constant depending on the structure}, we also examined the Si concentration dependence of the exchange stiffness constant for Fe-Al-Si with a DO3 structure.
The results are shown in Fig. \ref{ADO3dep}. We can see that if we increase the Si concentration, the exchange stiffness constants also increase.
{A} similar trend is also confirmed in {the}  A2 structure; however, the change {in}  amplitude of the exchange stiffness constant is small.
From these results, we can conclude that doping of {the}  Al atom leads to {a} reduction of the exchange stiffness constant for all structures {but} this trend {is most prominent in the B2 structure}. Although we took specific alloys in this case, {a} similar tendency is also experimentally confirmed in Co based Heusler alloys\cite{Kubota2009}.
In $\textrm{Co}_{2}\textrm{Mn}\textrm{Si}$ and $\textrm{Co}_2\textrm{Mn}\textrm{Al}$, $A=14.7\ \textrm{meV}/$$\textrm{\AA}$ and  $A=2.99\ \textrm{meV}/$$\textrm{\AA}$, respectively. However, the observed Curie temperatures are $985$ K and $693$ K. The difference between alloys is small compared with that of the exchange stiffness constant, and this tendency is qualitatively consistent with our results. 
\\
\section{Summary} 
In summary, we investigated the magnetic properties of Fe-Al-Si alloy not only at zero-temperature but also at finite temperatures.
{We confirmed the itinerant characteristics of Fe in Fe-Al-Si alloys.}
We also confirmed that {the} calculated exchange stiffness constant of {the} B2 structure is significantly small compared with that of other structures. However, calculated Curie temperatures are in {the} same order for all structure {that} are assumed in this work. 
In addition, we found that {the Fe atoms {that}  are closest to the Al and Si atoms exhibit smaller magnetic moments than other Fe atoms; this may reflect {the}  itineracy of these fe atoms owing to the strong hybridization with the $2p$ orbitals of Al and Si atoms.}
Moreover, {the} calculated temperature dependence of {the}  exchange stiffness constant of {the} B2 structure is also unique; this might be related to the spin configuration dependence of the exchange interactions, which deviates from the behavior of the usual Heisenberg model.  We also proposed that Si doping can increase the amplitude of the exchange stiffness constant.
\\
\ 
\begin{acknowledgments}
This work was supported by JSPS KAKENHI Grant Numbers  JP19H05612 in Japan.
\end{acknowledgments}

\end{document}